\documentclass{rnaastex}

\begin{document}

\title{Explaining the elongated shape of 'Oumuamua by the Eikonal abrasion model}

\correspondingauthor{Gyula M. Szab\'o}
\email{szgy@gothard.hu}

\author{G\'abor Domokos}
\affiliation{Dept. of Mechanics, Materials and Structures, Budapest University of Technology,
        M\H uegyetem rakpart 1-3., Budapest, Hungary, 1111}
\affiliation{MTA-BME Morphodynamics Research Group}
\email{domokos@iit.bme.hu}
\author{Andr\'as \'A. Sipos}
\affiliation{Dept. of Mechanics, Materials and Structures, Budapest University of Technology,
        M\H uegyetem rakpart 1-3., Budapest, Hungary, 1111}
\affiliation{MTA-BME Morphodynamics Research Group}
\email{siposa@eik.bme.hu}
\author{Gyula M. Szab\'o}
\affiliation{ELTE E\"otv\"os Lor\'and University, Gothard Astrophysical Observatory, Szombathely, Szent Imre h. u. 112, Hungary, 9700}
\author{P\'eter L. V\'arkonyi}
\affiliation{Dept. of Mechanics, Materials and Structures, Budapest University of Technology,
        M\H uegyetem rakpart 1-3., Budapest, Hungary, 1111}
\email{vpeter@mit.bme.hu}


\section{Introduction}

The photometry of the minor body with extrasolar origin (1I/2017 U1) 'Oumuamua revealed an unprecedented shape: \cite{2017Natur} reported a shape elongation b/a close to 1/10, which calls for theoretical explanation. Here we show that the abrasion of a primordial asteroid by a huge number of tiny particles ultimately leads to such elongated shape. The model (called the Eikonal equation) predicting this outcome was already suggested in \cite{2009ApJ...699..L13D} to play an important role in the evolution of asteroid shapes.

Disruptive collisions (e.g. among asteroids) generate primordial fragments with average axis ratios $2:\sqrt{2}:1$ (\cite{2000AREPS...28..367R,2008Icar..196..135S,2015NatSR...5..9147}). 
Despite substantial variation, extreme ratios close to $1:10$ have never been observed thus Oumuamua is very unlikely to be primordial fragment.

The primordial fragment then starts evolving via non-disruptive impacts, where the outcome is primarily determined by the fragment/impactor mass ratio, $M/m$. These impacts may include mergers (which may have shaped several Solar System asteroids), nevertheless, mergers can hardly explain elongations beyond 1:3. Large to moderate mass ratio $m\geq M$ lead to curvature-driven abrasion (\cite{2014PloSO...88657}) which tends to make objects rounder. Another class of low-energy impact collisions with high speed impactors of mass $m<<M$ has opposite results: it makes the asteroid less spherical.

\section{The model}

\begin{figure}
\begin{center}
\includegraphics[width=12cm]{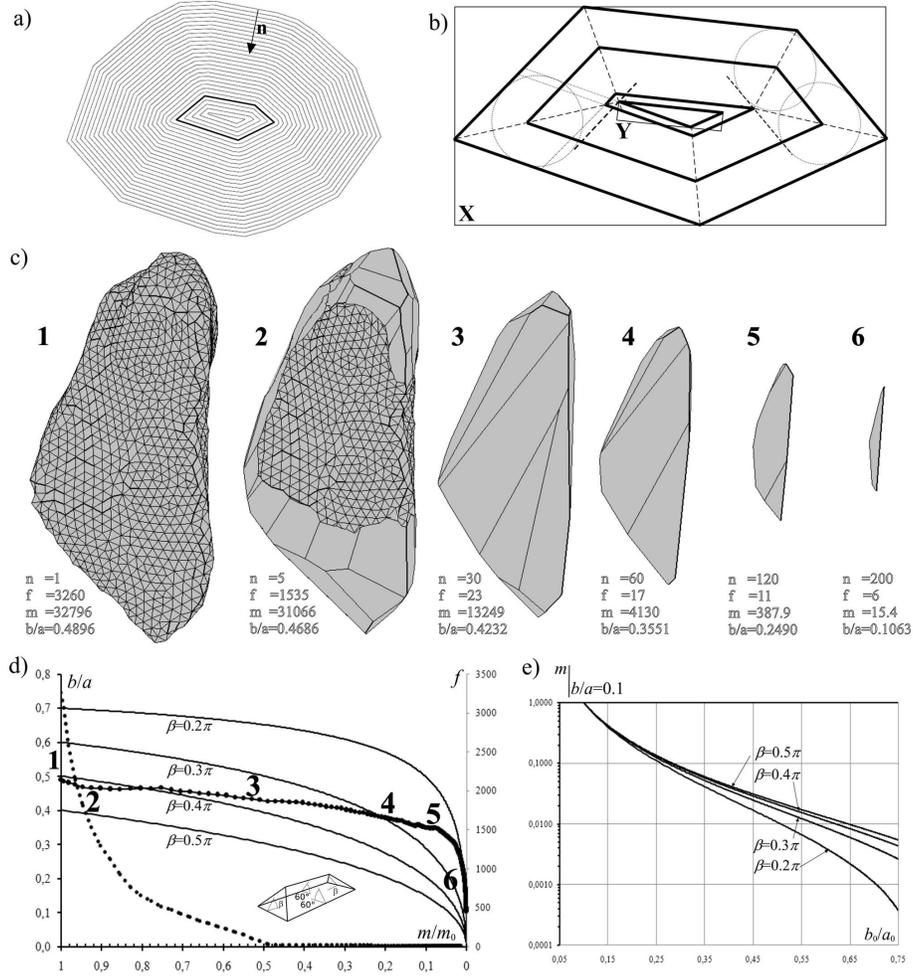}
\caption{(a) Eikonal equation acting on a planar polygon: observe gradual decrease of the number of edges. (b) Last stage of the same evolution, observe rapid change of elongation. (c) Simulation of Eikonal evolution of scanned fragment. $n$=number of collisions, $f$=number of faces, $m$=mass, $b/a$=elongation. (d) Elongation vs relative mass loss for previous fragment and some polyhedra. Observe that latter bracket the evolution of the former. Observe gradual decrease of $f$ and rapid decrease of $b/a$ in the second phase. (e) Relative mass at $b/a=1/10$ vs initial value of $b/a$.
\label{fig:1}}
\end{center}
\end{figure}

Very small abrading dust particles can be described by the limit $m/M\to 0$ where we have an exact geometric model, the so-called Eikonal equation, describing the evolution of the shape of the main body. Eikonal abrasion evolves shapes \emph{away from the sphere}  and shows the following generic, global properties:

(A) Large flat areas and sharp edges emerge spontaneously as dominant geometric features. This feature is strongly apparent already in the initial phase of Eikonal abrasion. If the initial shape is approximated by a multi-faceted polyhedron then the number $f$ of faces drops quickly until it reaches its ultimate value of $f=3$ or $f=4$.

(B) Shape elongation is dramatically increased during the evolution. In the initial phase of the evolution elongation is varying slowly. In the ultimate abrasion phase, where the mass of the object is a small fraction of the initial mass, the change is rapid and extreme elongation is reached. The elongation of the primordial shape is primarily influencing \emph{when} (i.e. at what mass loss percentage) extreme elongation is reached and it has much less influence on its actual value.

In our paper (\cite{2009ApJ...699..L13D}) we discussed the early stages of abrasion on asteroids where property (A)
is dominant because the asteroids observed in the Solar System are likely in the first phase of this evolution process.
We suggested that they suffer disruptive and/or merger collisions with large bodies frequently enough, and their global shape is regularly restructured and thus their evolution never reaches the end stages of Eikonal abrasion. Now we appeal to property (B) because 'Oumuamua may well have already reached this second, ultimate phase. Most likely it had sufficient time to evolve far from its original primordial shape and its current geometry may be interpreted as a mature, far evolved outcome of slow abrasion by tiny particles. 

\section{Results}

In Fig. 1 we present abrasion scenarios that have led from a  variety of initial shapes via the Eikonal evolution to highly elongated shapes. Observe the geometric features (A) and (B) on all examples.

'Oumuamua is the first known astronomical object with such extreme shape elongation. In its case, a long, uninterrupted slow abrasion is plausible, since the body resided in the interstellar space for several hundred million years (\cite{2017Natur}). Being free from larger impactors, the body was traveling at approximately 50 km/s velocity, passed through two solar systems, and suffered the impacts from many micrometer-sized interstellar dust grains with large velocities (in the order of the velocity dispersion of the Galactic thin disk, several 10 km/s). Collisions with small grains were energetic enough to dislodge splinters from the main body. A further evidence for an abrasion history is the presumedly complete lack of dust on the surface (\cite{2017Natur}).
Assuming an abrasion rate of 2--5$\mu$m/yr, the abrasion process could have easily evolved to the currently observed shape, regardless of the initial form.

A quantitative interpretation would have to rely on largely unknown data, such as the solar system where 'Oumuamua was formed, its initial form, cosmic path and internal composition. However, this example led to the recognition of the qualitative significance of the Eikonal abrasion. Further work is needed to establish quantitative evidence. 

\acknowledgments
The authors acknowledge the support NKFIH grants GINOP-2.3.2-15-2016-00003, K-119245 and K-125015.

\end{document}